\documentclass[journal]{IEEEtran}

\usepackage{amsmath,amsfonts,amssymb,dsfont}
\usepackage{epsfig}
\usepackage{amsfonts}
\usepackage{graphics}
\usepackage{graphicx}
\usepackage{subfig}
\usepackage{nopageno}

\epsfverbosetrue

\newcommand{\qed}{\hfill \ensuremath{\Box} \\}
\newcommand{\pf}{{\noindent \bf Proof: \ }}
\newcommand{\rk}{{\noindent \bf Remark: \ }}

\title{Network Capacity Region of Multi-Queue Multi-Server Queueing System with Time Varying Connectivities } 

\author{Hassaan Halabian, Ioannis Lambadaris, Chung-Horng Lung\\
Department of Systems and Computer Engineering\\
Carleton University, 1125 Colonel By Drive, Ottawa, ON, K1S 5B6 Canada\\
Email: \{hassanh, ioannis.lambadaris, chung-horng.lung\}@sce.carleton.ca\ 
} 

\begin{document}  
\maketitle   
\newtheorem{theor}{Theorem}
\newtheorem{lem}{Lemma}
\newtheorem{pro}{Proposition}
\newtheorem{defin}{Definition}
\newtheorem{conj}{Conjecture}
\newtheorem{cor}{Corollary}


\thispagestyle{empty}
\begin{abstract}
Network capacity region of multi-queue multi-server queueing system with random ON-OFF connectivities and stationary arrival processes is derived in this paper. Specifically, the necessary and sufficient conditions for the stability of the system are derived under general arrival processes with finite first and second moments. In the case of stationary arrival processes, these conditions establish the network capacity region of the system. It is also shown that AS/LCQ (Any Server/Longest Connected Queue) policy stabilizes the system when it is stabilizable. Furthermore, an upper bound for the average queue occupancy is derived for this policy. 
\end{abstract}
\footnotetext[1]{This work was supported by Mathematics of Information Technology and Complex Systems (MITACS) and Natural Sciences and Engineering Research Council of Canada (NSERC).}
\section {Introduction}
Resource allocation is one of the main concerns in the design process of emerging wireless networks. Examples of such networks are OFDMA and CDMA wireless systems in which orthogonal resources (OFDM subcarriers and CDMA codes) must be allocated to multiple users. Research in this area focuses on finding optimal policies to allocate orthogonal subchannels to the users. There are stochastic arrivals for each user which may be buffered to be transmitted in the future. Therefore, the resource allocation problem can be modeled as a multi-queue multi-server queueing system with parallel queues competing for available servers (which may model orthogonal subchannels \cite{javidi,javidi2,javidi3,javidi4, hussein, ganti}). However, because of users mobility, environmental changes, fading and etc., connectivity of each queue to each server is changing with time randomly. Thus, we are faced to a multi-queue multi-server system with time varying channel quality for which we have to design an appropriate server allocation policy. One of the main performance attributes which must be considered for each policy is its capacity region and how much this region coincides with the \emph{network capacity region} \cite{Now}. The capacity region of a network is defined as the closure of the set of all arrival rate matrices for which there exists an appropriate policy that stabilizes the system \cite{Now}. This region is unique for each network and is independent of resource allocation policy. 
On the other hand, the capacity region of a specified policy, say \(\pi\), is the closure of the set of all arrival rate matrices for which \(\pi\) results into the stability of the system. Obviously, the capacity region of any policy is a subset of the network capacity region. In fact, the network capacity region of a system is the union of the capacity regions of all the possible resource allocation policies we can have for a network \cite{Now}. A policy that achieves the network capacity region is called \textit{throughput optimal}.

The stability problem in wireless queueing networks was mainly addressed in \cite{Tassiulas92, Tassiulas93,Now, phd}. In \cite{Tassiulas92}, authors introduced the capacity region of a queueing network. They considered a time slotted system in their work and assumed that arrival processes are i.i.d. sequences and the queue length process is a Markov process. They also characterized the network capacity region of multi-queue single-server system with time varying ON-OFF connectivities which is described by some conditions on the arrival traffic \cite{Tassiulas93}. They also proved that for a symmetric system (with the same arrival and connectivity statistics for all the queues), LCQ (Longest Connected Queue) policy maximizes the capacity region and also provides the optimal performance in terms of average queue occupancy (or equivalently average delay) \cite{Tassiulas93}. In \cite{Now,phd} and \cite{power}, the notion of network capacity region of a wireless network was introduced for more general arrival and queue length processes. Furthermore, Lyapunov drift techniques were applied in \cite{Now} and \cite{phd} to analyse the stability of the proposed policies for stochastic optimization problems in wireless networks.

The problem of server allocation in multi-queue multi-server systems with time varying connectivities was mainly addressed in \cite{javidi,javidi2,javidi3,javidi4, hussein}. In \cite{javidi2}, Maximum Weight (MW) policy, a throughput optimal server allocation policy for stationary connectivity processes was proposed. However, \cite{javidi2} does not explicitly mention the conditions on the arrival traffic to guarantee the stability of MW. References \cite{javidi,javidi3,javidi4,hussein} study the optimal server allocation problem in terms of average delay. In \cite{javidi,javidi3,javidi4}, authors argue that in general, achieving instantaneous throughput and load balancing is impossible in a policy. However, as they show this goal is attainable in the special case of ON-OFF connectivity processes. They also introduced the MTLB (Maximum-Throughput Load-Balancing) policy and showed that this policy is minimizing a class of cost functions including total average delay for the case of two symmetric queues (with the same arrival and connectivities statistics). \cite{hussein} considers this problem for general number of symmetric queues and servers. Authors in \cite{hussein} characterized a class of \textit{Most Balancing} (MB) policies among all work conserving policies which are minimizing a class of cost functions including total average delay in stochastic ordering sense. They used stochastic ordering and dynamic coupling arguments to show the optimality of MB policies for symmetric systems.

In this paper, we will characterize the capacity region of multi-queue multi-server queueing system with random ON-OFF connectivities and stationary arrival processes based on the stochastic properties of the system. Toward this, the necessary and sufficient conditions for the stability of the system is derived under a general arrival process with finite first and second moments. For stationary arrival processes, these conditions establish the network capacity region of the system. We also showed that a simple server allocation policy called AS/LCQ maximizes the capacity region i.e. its capacity region coincide with the network capacity region and therefore it is a throughput optimal policy. It is worth mentioning that AS/LCQ acts exactly the same as MW policy proposed in \cite{javidi2} when the connectivity process is ON-OFF in MW.

The rest of the paper is organized as follows. Section II describes the model and notation required through the paper. In section III we discuss about the \textit{strong stability} definition in queueing networks and Lyapunov drift technique briefly. Then, we will derive necessary and sufficient conditions for the stability of our model and also find an upper bound for the average queue occupancy. In section IV we present simulation results and compare stability and delay performances of some heuristic work-conserving policies with those of AS/LCQ and the upper bound obtained in section III. Section V summarizes the conclusions of the paper.

\section {Model Description}

\begin{figure}[h!]
    \centering
    \includegraphics[width=0.45\textwidth]{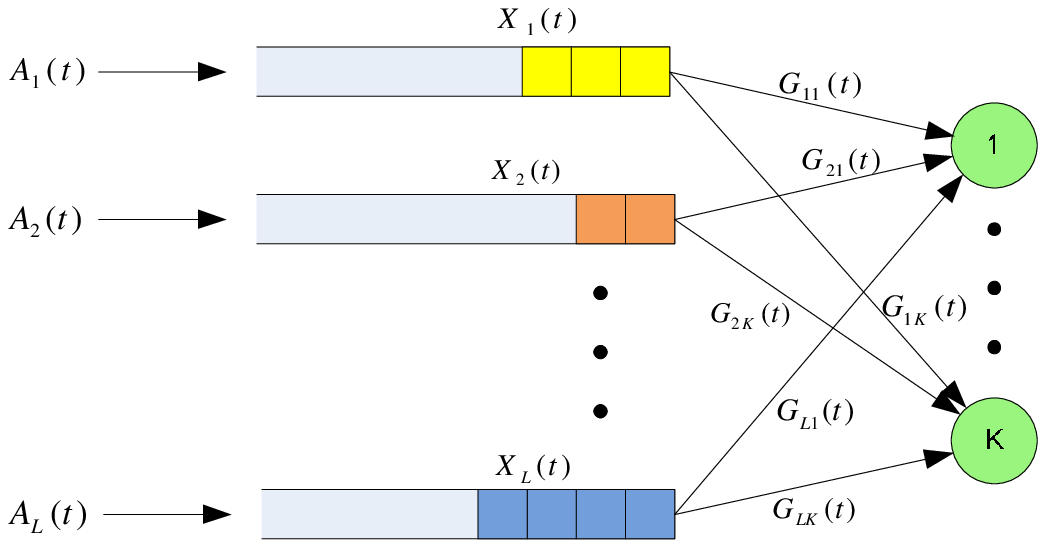}
	\caption{ Multi-queue multi-server queueing system with time varying connectivities}
	\label{model}
\end{figure}

Our model in this paper is the same as the model used in \cite{javidi,javidi2,javidi3,javidi4,hussein} with ON-OFF connectivity processes. We consider a time slotted queueing system with equal length time slots and equal length packets. The model consists of a set of parallel queues \(\cal L\) and a set of identical servers \(\cal K\). Each server can serve at most one packet at each time slot and we do not allow server sharing by the queues. In other words, each server can serve at most one queue at each time slot. Assume that \( \left|{\cal L}\right| = L \)  and  \( \left|{\cal K}\right| = K \). At each time slot \(t\), the link between each queue \( i\in \{ {1,...,L}\}\) and server \(j\in \{ {1,...,K}\}\) is either connected or disconnected. Assume that connectivity process between queue \(i\) and server \(j\) is modelled by an i.i.d. binary random process which is denoted by \(G_{ij}(t)\), i.e. \(G_{ij}(t)\in \{0,1\}\). Suppose that \(p_{ij}\) represents  the expected value of this process, i.e. \(E[G_{ij}(t)]=p_{ij}\). There are also exogenous arrival processes to the queues in set \(\cal L\). Assume that the arrival process to each queue \(i\) at time slot \(t\) (i.e. the number of packet arrivals during time slot \(t\)) is represented by \(A_i(t)\). For these processes we assume that \(E[A_i^2(t)]< A_{max}^2<\infty\) for all \(t\). Each queue has an infinite buffer space i.e. we do not have packet drops. We assume that the new arrivals are added to each queue at the end of each time slot. Let \(X(t)=(X_1(t), ... , X_L(t))\) be the queue length process vector at the end of time slot \(t\) after adding new arrivals to the queues. Figure \ref{model} shows the model used in this paper.

A server scheduling policy at each time slot should decide on how to allocate servers from set \(\cal K\) to the queues in set \(\cal L\). This must be accomplished based on the available information about the connectivities \(G_{ij}(t)\) and also the queue length process \(X(t)\). 
\section{Stability of Multi-Queue Multi-Server System with Time Varying Connectivities}
In this section, we will consider the stability problem of multi-queue multi-server system with ON-OFF connectivities for which we will find the necessary and sufficient conditions for its stability. We also show that AS/LCQ (Any Server/Longest Connected Queue) policy will stabilize the system as long as it is stabilizable. The details of this policy will be presented in part \textit{D} of this section. At first, we will have a review on the notion of strong stability in queueing networks.
\subsection{Strong Stability}     
We begin with introducing the definition of strong stability for a queueing system \cite{Now, phd}. Other definitions can be found in \cite{Asmussen, Tassiulas92, Tassiulas93, Leonardi}.
Consider a discrete time single queue system with an arrival process \(A(t)\) and service process \(\mu (t)\). Assume that the arrivals are added to the system at the end of each time slot. We can see that the queue length process \(X(t)\) at time \(t\) evolves with time according to the following rule.
\begin{equation}
X(t)=(X(t-1)-\mu(t))^+ +A(t)
\end{equation}
where \((\cdot)^+\) outputs the term inside the brackets if it is nonnegative and is zero otherwise.
Strong stability is given by the following definition \cite{Now}.
\begin{defin} 
A queue satisfying the conditions above is called \emph{strongly stable} if 
\begin{equation}
\limsup _{t \to \infty} {1\over{t}} \displaystyle\sum _ {\tau =0}^{t-1}E[X(\tau)] < \infty
\end{equation}   
\end{defin} 

Naturally for a queueing system we have the following definition \cite{Now}.
\begin{defin}  
A queueing system is called to be strongly stable if all the queues in the system are strongly stable. 
\end{defin} 

In our work we use the strong stability definition and from now we use ``stability" and ``strong stability" interchangably.

The following important property of strongly stable queues gives an invaluable insight of the above definitions.

\emph{Lemma 1} \cite{Now}:\label{zerolimit} If a queue is strongly stable and either \(E[A(t)] \leq A\) for all \(t\) or \(E[\mu (t)-A(t)] \leq D\) where \(A\) and \(D\) are finite nonnegative constants, then 
\begin{equation}
\lim _{t \to \infty} {1\over{t}} E[X(t)] =0
\end{equation}

A very important and useful mathematical tool used in network stability analysis and stochastic control/optimization of wireless networks is \emph{Lyapunov Drift} technique. We now present a brief review of this technique.
\subsection{Lyapunov Drift}
The basic idea behind the Lyapunov stability method is to define a nonnegative function of queue backlogs in a queueing system which can be seen as a \emph{measure} of the total aggregated backlog in the system at time \(t\). Then we evaluate the ``drift" of such function in two successive time slots by taking the effect of our control decision (scheduling or resource allocation policy) into account. If the expected value of the drift is negative as the backlog goes beyond a fixed threshold, then the system is stable. This is the method used in \cite{Now,phd,McKeown, Kumar,power,  Leonardi} to prove the stability of the systems working under their proposed policies.

For a queueing system with \(L\) queues and queue length vector \(X(t)=(X_1(t), ... , X_L(t))\), the following quadratic Lyapunov function has been used in literature (\cite{Now, phd, McKeown, power,Leonardi}).
\begin{equation}
\label{lypfunc}V(X)= \displaystyle\sum_{i=1}^L X_i^2(t)
\end{equation}
Assume that \(E[X_i(0)]<\infty\), \(\forall i=1, 2, ..., L\) and \(X(t)\) evolves with some probabilistic law (not necessarily Markovian). Then, the following important lemma holds.

\emph{Lemma 2} \cite{Now}:
If there exist constants \(B>0\) and \(\epsilon >0\) such that for all time slots \(t\) we have
\begin{eqnarray} 
 \label{lyapunov} E[V(X(t+1))-V(X(t)) \mid X(t)] \leq B - \epsilon\displaystyle\sum _{i=1}^L X_i(t),
\end{eqnarray}
then the system is strongly stable and further we have
\begin{equation}
\limsup_{t \to \infty} {1\over{t}} \displaystyle\sum_ {\tau =0}^{t-1} \displaystyle\sum _{i=1}^L E[X_i(\tau)] \leq \frac{B}{\epsilon}
\end{equation}
 
The left hand side of expression (\ref{lyapunov}) is usually called Lyapunov drift function which is a measure of expected value of changes in the backlog in two successive time slots. We can easily see the idea behind Lyapunov method in stabilizing queueing systems from Lemma 1. It is not hard to show that, when the aggregated backlog in the system goes beyond the bound \({B} \over \epsilon\), then the Lyapunov drift in the left hand side of (\ref{lyapunov}) will be negative, meaning that the system receives a negative drift on the expected aggregated backlog in two successive time slots. In other words the system tends toward lower backlogs and this results in its stability.

\subsection{Necessary Condition for the Stability of the System}
Let \(h_{ik}(t)\) be the departure process at time slot \(t\) from queue \(i\) to  server \(k\). Then, we can have the following equation for the queue length process which shows the evolution of queue length  process with time.
\begin{equation} 
\label{evol} X_i(t)=X_i(t-1)-\displaystyle\sum _{k=1}^K h_{ik}(t) +A_i(t)
\end{equation}
To find the necessary condition for the stability of the system, we need to use the following lemma. 

\emph{Lemma 3}: \label{equallimit}If the system is strongly stable under some server allocation policy \(\pi\), then for each queue \(i\)
\begin{eqnarray}
\label{mylem}\lim_{t\rightarrow \infty}\frac{1}{t}\displaystyle\sum_{\tau=1}^{t}E[A_i(\tau)]=\lim_{t\rightarrow \infty} \frac{1}{t}\displaystyle\sum_{\tau=1}^{t}\displaystyle\sum_{k=1}^K E[h_{ik}(\tau)],
\end{eqnarray}
i.e. for a stable system the average expected arrivals to a queue is equal to the average expected departure from that queue.
\pf See appendix A.

We now proceed to find the necessary condition for the stability of the system.
\begin{theor} \label{necessary} If there exists a server allocation policy \(\pi\) under which the system is stable, then 
\begin{eqnarray}
\label {theo1eq} \lim_{t\rightarrow\infty}\frac{1}{t}\displaystyle\sum_{\tau=1}^{t}\displaystyle\sum_{i\in Q}E[A_i(\tau)]\leq K-\displaystyle\sum_{k=1}^{K}\displaystyle\prod_{i\in Q}(1-p_{ik})\\
\nonumber \forall Q\subset \{1,...,L\}
\end{eqnarray}
\end{theor}
\pf See appendix B.

\rk If the arrival processes \(A_i(t)\)'s are stationary, then \(E[A_i(t)]= \lambda_i\) for all \(t\) and therefore the left hand side of (\ref{theo1eq}) will be equal to \(\sum_{i\in Q}\lambda_i\). Consequently, the necessary condition for the stability of the system with stationary arrival processes would be 
\begin{eqnarray}
\label{iid} \displaystyle\sum_{i\in Q}\lambda_i \leq \displaystyle\sum_{k=1}^{K}(1-\displaystyle\prod_{i \in Q}(1-p_{ik}))\;\;\;\;\; \forall Q\subset \{ {1,...,L}\}.
\end{eqnarray}
\subsection{Sufficient Condition for the Stability of the System}
 
We can divide the server allocation policy in our model into two scheduling problems.
First, we should determine the order under which servers are selected for service and second, for each server decide to allocate it to a particular queue. Consider the policy that chooses an arbitrary ordering of servers and then for each server, allocates it to its longest connected queue (LCQ). In other words, in this policy we do not restrict ourselves with a specific ordering of servers and we accept any permutation of the servers according to which servers will be selected for service. However, for the next phase of scheduling, for each selected server we use the LCQ policy. We call such a policy as AS/LCQ (Any Server/Longest Connected Queue). We will now derive the sufficient condition for the stability of our model and prove that AS/LCQ stabilizes the system as long as condition (\ref{sc1}) is satisfied. An upper bound is also derived for the time averaged expected number of packets in the system.

\begin{theor}\label{sc} The multi-queue multi-server system is stable under AS/LCQ if \emph{for all} \(t\)
\begin{equation}
\label{sc1}\displaystyle\sum_{i\in Q} E[A_i(t)] < K-\displaystyle\sum_{k=1}^{K}\displaystyle\prod_{i\in Q}(1-p_{ik}) \;\;\; \forall Q\subset \{ {1,...,L}\}.
\end{equation}
Furthermore, the following bound for the average expected ``aggregate" occupancy holds.
\begin{eqnarray}
\label{sc2} \lefteqn{\limsup_{t \to \infty} {1\over{t}} \displaystyle\sum_ {\tau =0}^{t-1} \displaystyle\sum_{i=1}^L E[X_i(\tau)]\leq} \\
&& \frac{ -\frac{L}{2} \left(LA^2_{max}+K(2K-1)\right)}
{\displaystyle\max_{Q\subset\{{1,2,...,L}\},t}\left\{\displaystyle\sum_{i \in Q} E[A_{i}(t)]- K +\displaystyle\sum_{k=1}^K \displaystyle\prod_{i\in Q}(1-p_{ik})\right\}}\nonumber
\end{eqnarray}
\end{theor}
\pf See appendix C.

It is worth mentioning that AS/LCQ acts exactly the same as MW policy proposed in \cite{javidi2} when the connectivity process is ON-OFF in MW. 

Note that for all the servers we only use the backlog information at the beginning of each time slot, i.e. during the implementation of AS/LCQ policy at each time slot we do not update the queue lengths until all the servers are allocated at which point we update the queue lengths. It is interesting to note that this policy can be non-work conserving at some time slots. In other words, there may exist some idle servers at a time slot while they could have served other backlogged queues. We will discuss about it through an example in part \textit{E} of this section.

\rk Note that by considering stationary assumption on the arrival processes, the condition (\ref{sc1}) would be
\begin{equation}
 \label{sc21}\displaystyle\sum_{i\in Q} \lambda _i < K-\displaystyle\sum_{k=1}^{K}\displaystyle\prod_{i\in Q}(1-p_{ik}) \;\;\;\;\;\;\;\;\; \forall Q\subset \{ {1,...,L}\}.
\end{equation}
According to the definition of system capacity region and (\ref{iid}) and (\ref{sc21}), equation (\ref{iid}) characterizes the network capacity region of multi-queue multi-server system with stationary ON-OFF connectivities and stationary arrivals.
\subsection{Discussion}
As mentioned earlier, AS/LCQ may exhibit non-work conserving behavior during some time slots. This can be clarified by the following example.

Consider a system with \(L=2\) and \(K=3\) with queue length vector \(X(t)=(2,1)\) at time slot \(t\). For the connectivities at this time we have the following matrix.
\[G(t)=\left[\begin{array}{ccc}
1 & 1 & 1 \\
0 & 0 & 1\end{array}\right]\]  

Assume that the ordering of server selection is server 1 first and then server 2 and finally server 3. Servers 1 and 2 both are allocated to queue 1 according to LCQ rule. Server 3 is connected to both of the queues. Since in AS/LCQ all the servers are allocated first and then the queue lengths are updated afterwards, queue 1 is the longest connected queue for server 3. Thus, server 3 is allocated to queue 1 as well. However, queue 1 has only two packets waiting for service and therefore server 3 will be idle at this time slot (although it could have been used to serve queue 2). 

Note that since AS/LCQ is a non-work conserving policy it can not be delay optimal. However, it can achieve the network capacity region as explained previously in part \textit{D}. In fact, AS/LCQ will exhibit non-work conserving behaviour in light arrival loads and as the load increases its behaviour will converge to work conserving. Since the capacity region of a system is mainly determined by its behaviour in heavy arrival loads, this property of AS/LCQ does not have conflict with its throughput optimality. 
It is worth mentioning that not all work-conserving policies are throughput optimal. In the following section by simulations we will observe that some work-conserving policies cannot achieve the network capacity region. In the following section, we will also observe that how the service ordering of servers affects the average total queue occupancy. However, as we showed in the previous part an arbitrary ordering is sufficient to achieve the network capacity region.

\section{Simulation Results}
Simulation is used to show the validity of our analysis in the previous section and also to compare performance of AS/LCQ to some heuristic work conserving policies including LCSF/LCQ (Least Connected Server First/Longest Connected Queue), MCSF/LCQ (Most Connected Server First/Longest Connected Queue), LCSF/SCQ (Least Connected Server First/Shortest Connected Queue), MCSF/SCQ (Most Connected Server First/Shortest Connected Queue) and a Randomized policy \cite{hussein}.

The LCSF (MCSF) policy at the first phase of scheduling (i.e. determination of servers order) will sort the servers for service according to their number of connectivities in an ascending (descending) order. The LCQ (SCQ) policy will assign the selected server it to its longest connected queue (shortest connected queue). Note that in order to make SCQ policies work conserving, we only serve the shortest non-empty queues. The Randomized policy at each time slot makes random server selections and for each server random non-empty queue selection.

\begin{figure}[h!]
    \centering
    \includegraphics[width=0.5\textwidth]{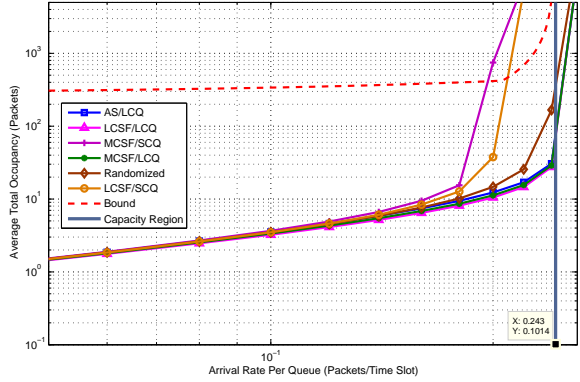}
	\caption{Average Total Occupancy for \(p=0.2\)}
	\label{fig2}
\end{figure}

\begin{figure}[h!]
    \centering
    \includegraphics[width=0.5\textwidth]{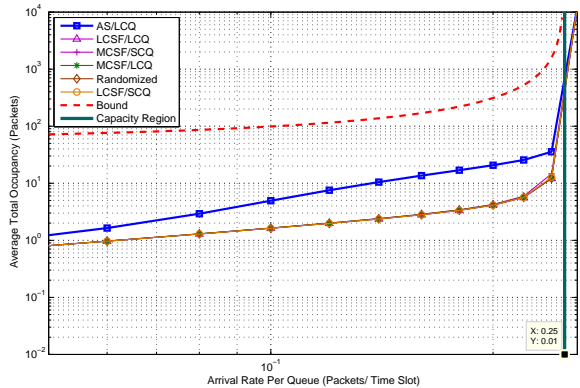}
	\caption{Average Total Occupancy for \(p=0.9\)}
	\label{fig9}
\end{figure}

We have simulated a system consisting of 16 queues (\(L=16\)) and 4 servers (\(K=4\)). First, we considered a symmetric system in which all the arrivals to all the queues are the same in distribution. We also assumed that connectivity variables have the same distribution (the same connectivity probabilities). In this system, arrivals are assumed to have i.i.d. Bernoulli distributions. The capacity region for these special cases would be an \(n\) dimensional cube whose side size is equal to \(\frac{K-K(1-p)^L}{L}\) which is 0.243 for \(p= 0.2\) and is almost 0.25 for \(p= 0.9\). Figures (\ref{fig2}) and (\ref{fig9}) show the average total occupancy of different policies for connectivity probabilities 0.2, and 0.9 versus arrival rate per queue. In these figures, it is observed that in all the cases if the arrivals are inside the capacity region, AS/LCQ can stabilize the system and has average total occupancy below the bound we derived in the previous section. 

We can further conclude that as the connectivity variable increases the performance of the work conserving policies become the same. This agrees with intuition since when the system is close to full connectivity, any work conserving algorithm will be optimal in terms of average occupancy and of course better than any non-work conserving policy like AS/LCQ. Although AS/LCQ has larger average total occupancy compared to other policies, it still stabilizes the system as long as arrivals are inside the capacity region and has bounded average total occupancy. However, this is not the case for LCSF/SCQ and MCSF/SCQ policies and they cannot stabilize the system for certain arrivals inside the capacity region. From these figures we also see that randomized policy performs very close to the other policies in these special cases and this is due to existence of symmetry (in arrivals and connectivities) in these cases.

\begin{figure}[h!]
    \centering
    \includegraphics[width=0.5\textwidth]{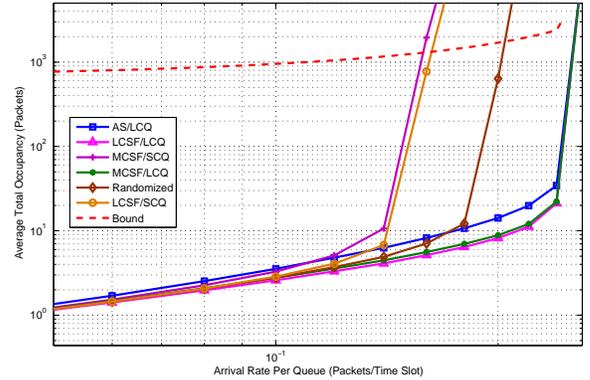}
	\caption{Average total occupancy for an asymmetric scenario}
	\label{asymmetric}
\end{figure}

We have also simulated an asymmetric system in which connectivity variables comes from the following matrix in which \(p_{ij}=E[G_{ij}(t)]\). This matrix was chosen randomly.
\[\hspace{-.5cm}p=\left(\begin{array}{lllllllllllllllllllll}
0.9 & 0.2 & 0.2 & 0.8 & 0.2 & 0.1 & 0.5 & 0.6 \\
0.8 & 0.1 & 0.1 & 0.9 & 0.02 & 0.5 & 0.8 & 0.8\\
0.9 & 0.02 & 0.5 & 0.99 & 0.3 & 0.8 & 0.78 & 0.99 \\
0.8 & 0.03 & 0.9 & 0.87 & 0.5 & 0.98 & 0.62 & 0.4 \end{array}\right.\]  
\[\left.\begin{array}{rrrrrrrrrrrrrrrrrrrrrrrrr}
 \;\;\;\;\;\; 0.2 & 0.72 & 0.86 & 0.3 & 0.66 & 0.21 & 0.84 & 0.03 \\
\;\;\;\;\;\;0.1 & 0.65 & 0.65 & 0.15 & 0.58 & 0.32 & 0.69 & 0.12 \\
\;\;\;\;\;\; 0.02 & 0.42 & 0.94 & 0.35 & 0.9 & 0.16 & 0.96 & 0.21 \\
\;\;\;\;\;\; 0.8 & 1 & 0.7 & 0.09 & 0.1 & 0.45 & 0.13 & 0.07 \end{array}\right)^T\] 
In this experiment, arrivals are following the Poisson distribution with the same rates for each slot and  each queue. The capacity region in this case is not easy to characterize and describe in a concise manner. Figure (\ref{asymmetric}) shows the results for this case. In this figure, we observe that Randomized policy could not capture the capacity region wholly. However, LCQ policies (AS/LCQ, LCSF/LCQ and MCSF/LCQ) performs similarly to each other from stability point of view.

From the above simulations, we can also observe that AS/LCQ performs slightly worse as compared with other policies in light arrival loads. This behaviour is because the fact that AS/LCQ may exhibit non-work conserving behaviour more frequently for light arrival loads. However, as the load increases AS/LCQ will be work conserving with high probability. We can also observe that the obtained bound is not tight.

\section{Conclusions}
In this paper we derived the necessary and sufficient conditions for the stability of multi-queue multi-server system with random connectivities and characterized the capacity region of this system for stationary arrivals. We also introduced AS/LCQ policy and argued that although this policy is a non-work conserving policy, it can stabilize the system for all the arrivals inside the capacity region and therefore it is a throughput optimal policy. Then, we derived an upper bound of the average queue occupancy for this policy. Finally, we used simulations to validate our analysis and compare this policy to some work conserving policies in terms of average queue occupancy.

If we modify the policy such that the queue lengths are updated after each server is allocated, we can establish a work conserving policy.
However, this does not increase the capacity region for a system with stationary arrivals. However, we may help us to obtain a tighter bound than we obtained in this work.

\appendices
\section{Proof of Lemma 3} 
\pf
If we write equation (\ref{evol}) for \(\tau=1,2,...,t\) and then adding them up, we will have
\begin{equation}
X_i(t)=X_i(0)-\displaystyle\sum _{\tau=1}^t \displaystyle\sum_{k=1}^K h_{ik}(t) + \displaystyle\sum _{\tau=1}^t A_i(t)
\end{equation}
Taking the expectation from both sides, dividing by \(t\) and then taking the limit as \(t\) goes to infinity, we will have the following.
\begin{eqnarray}
\lim_{t\rightarrow \infty} \frac{E[X_i(t)]}{t}=\lim_{t\rightarrow \infty} \frac{E[X_i(0)]}{t} \nonumber \\ -  \lim_{t\rightarrow \infty} \frac{1}{t} \displaystyle\sum _{\tau=1}^t \displaystyle\sum_{k=1}^K E[h_{ik}(t)]   + \lim_{t\rightarrow \infty} \frac{1}{t} \displaystyle\sum _{\tau=1}^t E[A_i(t)] 
\end{eqnarray}
According to Lemma 1 and the assumption that \(E[X_i(0)]<\infty\), the left hand side term and the first term in the right hand side term are equal to zero and therefore the result is proven.
\qed

\section{Proof of Theorem 1} 
Since the system is strongly stable, (\ref{mylem}) must be satisfied for any subset of queues \(Q\subset\{{1,...,L}\}\), i.e. 
\begin{eqnarray}  
\label{t1l1}\lim_{t\rightarrow \infty}\frac{1}{t}\displaystyle\sum_{\tau=1}^{t}\displaystyle\sum_{i\in Q}E[A_i(\tau)]=\lim_{t\rightarrow\infty}\frac{1}{t}\displaystyle\sum_{\tau=1}^{t}\displaystyle\sum_{i\in Q}\displaystyle\sum_{k=1}^K E[h_{ik}(\tau)]
\end{eqnarray}
We now define the sets \(B_k(\tau)\) as
\begin{equation} 
B_k(\tau)=\{ G_{\ell k}(\tau),X_{\ell}(\tau-1), \ell \in Q\}
\end{equation}
For \(B_k(\tau)\), three disjoint cases are imaginable. 

\(B_k^1(\tau)=\{ G_{ik}(\tau)=0,i \in Q\}\)

\(B_k^2(\tau)=\{ G_{ik}(\tau)=0, i \in Q\}^c\cap \{X_{i}(\tau-1)=0, i \in Q \}\)

\(B_k^3(\tau)=\{ G_{ik}(\tau)=0, i \in Q\}^c\cap \{X_{i}(\tau-1)=0, i \in Q \}^c\)
By conditioning each term in the right hand side summation in (\ref{t1l1}) to the event \(B_k(\tau)\) we have 
\begin{equation} 
\label{t1l2}\displaystyle\sum_{k=1}^{K}\displaystyle\sum_{i\in Q}E[h_{ik}(\tau)]=\displaystyle\sum_{k=1}^{K}E_{B_k(\tau)}\left[E\left[\displaystyle\sum_{i\in Q}h_{ik}(\tau)\mid B_k(\tau) \right ]\right]
\end{equation}
We can easily see that  
\begin{equation} 
\label{t1l3}E\left [ \displaystyle\sum_{i\in Q}h_{ik}(\tau)\mid B_k^j(\tau) \right ]=0 \;\;\;\;\;\;\;\;\;\; j=1,2 
\end{equation}
and
\begin{equation}
\label{t1l4}E\left [ \displaystyle\sum_{i\in Q}h_{ik}(\tau)\mid B_k^3(\tau) \right ]\leq 1   
\end{equation} 
Using (\ref{t1l3}) and (\ref{t1l4}), equation (\ref{t1l2}) can be simplified to the following.
\begin{eqnarray}
\label{t1l5}\displaystyle\sum_{k=1}^{K}\displaystyle\sum_{i\in Q}E[h_{ik}(\tau)] \leq \displaystyle\sum_{k=1}^{K}(1-P[B_k^1(\tau)]-P[B_k^2(\tau)])
\end{eqnarray}
Note that \(P[B_k^2(\tau)]\geq 0\) and for \(P[B_k^1(\tau)]\), we have
\begin{equation} 
\label{t1l6}P[B_k^1(\tau)]=\displaystyle\prod_{i \in Q}(1-p_{ik})
\end{equation} 
Finally, from (\ref{t1l1}), (\ref{t1l5}) and (\ref{t1l6}) we conclude that
\begin{eqnarray}  
\lim_{t\rightarrow \infty}\frac{1}{t}\displaystyle\sum_{\tau=1}^{t}\displaystyle\sum_{i\in Q}E[A_i(\tau)] \leq \displaystyle\sum_{k=1}^{K}(1-\displaystyle\prod_{i \in Q}(1-p_{ik}))
\end{eqnarray}
and the theorem follows.
\qed

\section{Proof of Theorem \ref{sc} }
\pf We will start with the Lyapunov function evaluation. we will use the quadratic function (\ref{lypfunc}) as our Lyapunov function. The Lyapunov drift for two successive time slots has the following form.
\begin{eqnarray}
\lefteqn {E[V(X(t+1))-V(X(t)) \mid X(t))]} \nonumber \\ & &= E\left[\displaystyle\sum_{i=1}^L X_i^2(t+1)-X_i^2(t) \mid X(t)\right] \nonumber \\
& &=  E\left[\displaystyle\sum_{i=1}^L (X_i(t+1)-X_i(t))^2 \mid X(t)\right]\nonumber \\ \label{sc3}
& &+ {\;\;\;}2 E\left[\displaystyle\sum_{i=1}^L X_i(t)(X_i(t+1)-X_i(t))\mid X(t)\right]
\end{eqnarray}

For the the first term we have:
\begin{eqnarray}
\notag \lefteqn{E\left[\displaystyle\sum_{i=1}^L (X_i(t+1)-X_i(t))^2 \mid X(t)\right]}\\ \notag
& &= E\left[\displaystyle\sum_{i=1}^L (A_i(t+1)-\displaystyle\sum_{k=1}^K h_{ik}(t))^2 \mid X(t)\right]\\ \notag
& &= E\left[\displaystyle\sum_{i=1}^L A_i^2(t+1) \mid X(t)\right] \\ \nonumber & & -2E\left[\displaystyle\sum_{i=1}^L\displaystyle\sum_{k=1}^K A_i(t+1)h_{ik}(t) \mid X(t)\right]\\ \label{sc4} & & + \;\;\; E\left[\displaystyle\sum _{i=1}^L \left ( \displaystyle\sum_{k=1}^K h_{ik}(t) \right )^2 \mid X(t)\right]
\end{eqnarray}
Using the the fact that \(\displaystyle\sum_{k=1}^K h_{ik}(t) \geq 0\) we get the following inequality
\begin{eqnarray}
\label{sc5} & \displaystyle\sum _{i=1}^L \left ( \displaystyle\sum_{k=1}^K h_{ik}(t) \right )^2  \leq \left ( \displaystyle\sum_{i=1}^L\displaystyle\sum_{k=1}^K h_{ik}(t)\right)^2 \leq K^2 
\end{eqnarray}
Since \(\displaystyle\sum_{i=1}^L\displaystyle\sum_{k=1}^K A_i(t+1)h_{ik}(t) \geq 0\), the first term in (\ref{sc3}) can be bounded by 
\begin{eqnarray}
 \label {sc6} \lefteqn{E\left[\displaystyle\sum_{i=1}^L (X_i(t+1)-X_i(t))^2 \mid X(t)\right]} \nonumber \\ 
 & &\leq  \displaystyle\sum_{i=1}^L E[A^2_i(t+1)]+K^2
\end{eqnarray}
Now assume that we select the servers for service according to an arbitrary order \(s_1, s_2, ..., s_K\). Thus, for the second term in (\ref{sc3}) we have
\begin{eqnarray}
\label{sc7}\notag \lefteqn {E\left[\displaystyle\sum_{i=1}^L X_i(t)(X_i(t+1)-X_i(t))\mid X(t)\right]}\\ 
\notag &&= E\left[\displaystyle\sum_{i=1}^L X_i(t)(A_i(t+1)-\displaystyle\sum_{k=1}^K h_{ik}(t+1))\mid X(t)\right] \\
 && = E\left[\displaystyle\sum_{i=1}^L X_i(t)A_i(t+1)\mid X(t) \right] \nonumber \\
 & & -E\left[\displaystyle\sum_{i=1}^L\displaystyle\sum_{k=1}^K X_i(t)h_{is_k}(t+1)\mid X(t) \right]   
\end{eqnarray}
The first term in (\ref{sc7}) can be written as follows. 
\begin{eqnarray}
\label{sc8}E\left[\displaystyle\sum_{i=1}^L X_i(t)A_i(t+1)\mid X(t) \right]=\displaystyle\sum_{i=1}^L E[A_i(t+1)] X_i(t)
\end{eqnarray}
For the second term in (\ref{sc7}) we have  
\begin{eqnarray}
\label{sc9}\lefteqn{E\left[\displaystyle\sum_{i=1}^L\displaystyle\sum_{k=1}^K X_i(t)h_{is_k}(t+1)\mid X(t) \right]} \nonumber \\&& = \displaystyle\sum_{k=1}^K E\left[\displaystyle\sum_{i=1}^L X_i(t)h_{is_k}(t+1)\mid X(t) \right]
\end{eqnarray}

Now, we introduce the following notation. We sort the queue length process at time slot \(t\) in an ascending order \(X_{q_1}, X_{q_2}, ....,X_{q_L}\) , i.e.  \(X_{q_i}(t) \geq X_{q_{i-1}}(t)\) for all \(i=2,...,L\) and if \(X_{q_i}(t)=X_{q_{i-1}}(t)\), then \(q_i \geq q_{i-1}\). Furthermore, consider the following decomposition of the connectivity processes for each server \(k\). 
\begin{eqnarray}
\notag \lefteqn{D_0^k=\left\{ G_{ik}(t+1)=0, \; \;for \;\; all \;\; i \in \{{1,...,L}\} \right\}}
\\  \notag  \lefteqn{D_i^k=\{ G_{q_ik}(t+1)=1 , G_{q_\ell k}(t+1)=0,} \\ \nonumber  && \;\;\;\;\;\;\;\;\;\;\;\;\;\;\;\;\;\;\;\;\;for\;\; i < \ell \leq L \;\; and \;\;\;all\;\; i \in \{1,...,L\} \}
\end{eqnarray}
The probability of events \(D_i^k\) is given by
\begin{eqnarray}
\notag & P(D_0^k)= \displaystyle\prod_{i=1}^L(1-p_{ik})  \;\;\;,\;\;\; P(D_i^k)= p_{q_ik}\displaystyle\prod_{u=i+1}^L(1-p_{q_uk})
\end{eqnarray}

In the second term of equation (\ref{sc9}), each term in the summation can be rewritten as
\begin{eqnarray}
\label{sc10}\notag \lefteqn{E\left[\displaystyle\sum_{i=1}^L X_i(t)h_{is_k}(t+1)\mid X(t)\right]} \\ && 
\notag  = E\left[\displaystyle\sum_{i=1}^L X_{q_i}(t)h_{q_is_k}(t+1)\mid X(t) \right] \\
 & &= \displaystyle\sum_{l=0}^L E\left[\displaystyle\sum_{i=1}^L X_{q_i}(t)h_{q_is_k}(t+1)\mid  X(t),D_l^{s_k} \right]P(D_l^{s_k}) \nonumber \\
\end{eqnarray}
Note that 
\begin{equation}
\label{sc11}\nonumber E\left[\displaystyle\sum_{i=1}^L X_{q_i}(t)h_{q_is_k}(t+1)\mid X(t),D_l^{s_k} \right] \geq (X_{q_l}(t)-(k-1))^+.
\end{equation}
Therefore, equation (\ref{sc10}) can be bounded by 
\begin{eqnarray}
\label{sc12}\notag \lefteqn{E\left[\displaystyle\sum_{i=1}^L X_i(t)h_{is_k}(t+1)\mid X(t)\right]} \\
 & & \geq  \notag \displaystyle\sum_{l=1}^L \left ( X_{q_l}(t)-(k-1) \right) p_{q_ls_k}\displaystyle\prod_{u=l+1}^L (1-p_{q_js_k})\\ \nonumber
 & &= \displaystyle\sum_{l=1}^L X_{q_l}(t) p_{q_l s_k}\displaystyle\prod_{j=l+1}^L (1-p_{q_js_k})\nonumber \\ & & \;\;\;\ -(k-1)\displaystyle\sum_{l=1}^L p_{q_l s_k}\displaystyle\prod_{j=l+1}^L (1-p_{q_j s_k})
\end{eqnarray}
For the second term in (\ref{sc12}) we have
\begin{eqnarray}
\label{sc13}\lefteqn{(k-1)\displaystyle\sum_{l=1}^L p_{q_l s_k}\displaystyle\prod_{j=l+1}^L (1-p_{q_j s_k})} \nonumber \\ && =(k-1)\left(1-\displaystyle\prod_{j=1}^{L}(1-p_{q_j s_k})\right)
\end{eqnarray}
and for the first term 
\begin{eqnarray}
\label{sc14}\notag \lefteqn{\displaystyle\sum_{l=1}^L X_{q_l}(t) p_{q_l s_k}\displaystyle\prod_{j=l+1}^L (1-p_{q_j s_k})} \\ \notag && = \displaystyle\sum_{j=2}^L (X_{q_j}(t)-X_{q_{j-1}}(t)) \left(1-\displaystyle\prod_{l=j}^L (1-p_{q_l s_k})\right) \\
 && \;\;\; + X_{q_1}(t)\left(1-\displaystyle\prod_{j=1}^{L}(1-p_{q_j s_k})\right)
\end{eqnarray}
Equation (\ref{sc8}) also can be written as follows.
\begin{eqnarray}
\label{sc15}\notag \lefteqn{\displaystyle\sum_{i=1}^L E[A_i(t+1)] X_i(t)=\displaystyle\sum_{l=1}^L E[A_{q_l}(t+1)] X_{q_l}(t)}
\nonumber \\
&& = \displaystyle\sum_{j=2}^L (X_{q_j}(t)-X_{q_{j-1}}(t))\displaystyle\sum_{l=j}^L E[A_{q_l}(t+1)]\nonumber \\ && \;\;\; +X_{q_1}(t)\displaystyle\sum_{l=1}^L E[A_{q_l}(t+1)] 
\end{eqnarray}

Using equations (\ref{sc7})-(\ref{sc9}) and (\ref{sc12})-(\ref{sc15}) we have the following bound for the second term in (\ref{sc3}).
\begin{eqnarray}
\label{sc16}\notag \lefteqn{ E\left[\displaystyle\sum_{i=1}^L X_i(t)(X_i(t+1)-X_i(t))\mid X(t)\right]}\\ 
\notag &&\leq  \displaystyle\sum_{j=2}^L (X_{q_j}(t)-X_{q_{j-1}}(t))\displaystyle\sum_{l=j}^L E[A_{q_l}(t+1)]\nonumber \\
&& \;\;\;+X_{q_1}(t)\displaystyle\sum_{l=1}^L E[A_{q_l}(t+1)] \nonumber \\ 
\notag && \;\;\;\;-\displaystyle\sum_{j=2}^L (X_{q_j}(t)-X_{q_{j-1}}(t)) \displaystyle\sum_{k=1}^K \left(1-\displaystyle\prod_{l=j}^L(1-p_{q_l s_k})\right)  \\ 
&& \;\;\;\; \notag -X_{q_1}(t)\displaystyle\sum_{k=1}^K\left(1-\displaystyle\prod_{j=1}^{L}(1-p_{q_j s_k})\right)\nonumber \\
&&\;\;\;\;+ \displaystyle\sum_{k=1}^K (k-1)\left(1-\displaystyle\prod_{j=1}^{L}(1-p_{q_j s_k}) \right) \nonumber\\
\notag && \leq \displaystyle\sum_{j=2}^L (X_{q_j}(t)-X_{q_{j-1}}(t)) \nonumber \\
&& ~~~ \cdot\left( \displaystyle\sum_{l=j}^L E[A_{q_l}(t+1)] - \displaystyle\sum_{k=1}^K \left(1-\displaystyle\prod_{l=j}^L (1-p_{q_ls_k})\right)\right) \nonumber \\
&& \notag~~~ +X_{q_1}(t) \left(\displaystyle\sum_{l=1}^L E[A_{q_l}(t+1)] \right. \\
  && ~~~~~ \left. -\displaystyle\sum_{k=1}^K\left(1-\displaystyle\prod_{j=1}^{L}(1-p_{q_js_k})\right)\right) + \displaystyle\sum_{k=1}^K (k-1)
\end{eqnarray}
Now, define \(m\) as
\begin{eqnarray}
\nonumber \label{sc17}m= \max_{Q\subset\{{1,2,...,L}\},t}\left\{\displaystyle\sum_{i \in Q} E[A_{i}(t)]- K + \displaystyle\sum_{k=1}^K \displaystyle\prod_{i\in Q} (1-p_{ik})\right\}
\end{eqnarray}
Therefore equation (\ref{sc16}) can be bounded by the following.
\begin{eqnarray}
\label{sc18}\notag \lefteqn{E\left[\displaystyle\sum_{i=1}^L X_i(t)(X_i(t+1)-X_i(t))\mid X(t)\right]} \\ 
&& \leq \notag \displaystyle\sum_{j=2}^L (X_{q_j}(t)-X_{q_{j-1}}(t))m + X_{q_1}(t)m + \frac{K}{2}(K-1)\\
&& = X_{q_L}(t)m+\frac{K}{2}(K-1)
\end{eqnarray}
Putting all together and according to (\ref{sc3}), (\ref{sc6}) and (\ref{sc18}) the Lyapunov drift in equation (\ref{sc3}) is upper bounded by the following.
\begin{eqnarray}
\label{sc19}\notag \lefteqn{E[V(X(t+1))-V(X(t)) \mid X(t))]} \\
&& \leq  \displaystyle\sum_{i=1}^L E[A^2_i(t+1)]+K^2+ 2 X_{s_L}(t)m \\ 
\notag && \;\;\; +K(K-1) \\ 
&&= \displaystyle\sum_{i=1}^L E[A^2_i(t+1)]+2K^2-K -\epsilon (LX_{s_L}(t)) 
\end{eqnarray}
where \(\epsilon=-\frac{2m}{L}\). According to condition (\ref{sc1}), \(m\) is negative, therefore \(\epsilon>0\). Since \(\sum_{i=1}^L X_i(t) \leq LX_{s_L}(t)\), therfore \(-\epsilon (LX_{s_L}(t)) \leq -\epsilon\sum_{i=1}^L X_i(t)\).
Consequently the Lyapunov drift (\ref{sc19}) is bounded by
\begin{eqnarray}
\label{sc20}\notag \lefteqn{E[V(X(t+1))-V(X(t)) \mid X(t))]}\\ 
&& \leq LA_{max}^2+2K^2-K -\epsilon\sum_{i=1}^L X_i(t) \nonumber\\
&& =B -\epsilon\sum_{i=1}^L X_i(t) 
\end{eqnarray}
in which \(B\) that has positive value is defined as 
\begin{eqnarray}
B=LA_{max}^2+2K^2-K
\end{eqnarray}
Therefore, according to Lemma 2, the multi-queue multi-server system is stable under AS/LCQ as long as condition (\ref{sc1}) is satisfied and also the time average expected congestion in the system is bounded by  
\begin{eqnarray}
 \limsup_{t \to \infty} {1\over{t}} \displaystyle\sum_ {\tau =0}^{t-1} \displaystyle\sum_{i=1}^L E[X_i(\tau)] \leq \frac{B}{\epsilon}
\end{eqnarray}
which is equal to (\ref{sc2}). 
\qed

\end{document}